\documentclass[twocolumn,showpacs,preprintnumbers,amsmath,amssymb,prl,epsf]{revtex4}
\usepackage{graphicx}

\begin{document}
\title{High-visibility multi-photon interference\\
of Hanbury Brown - Twiss type for classical light}
\author{I.~N.~Agafonov, M.~V.~Chekhova, T.~Sh.~Iskhakov, A.~N.~Penin}
\affiliation{Department of Physics, M.V.Lomonosov Moscow State
University,\\  Leninskie Gory, 119992 Moscow, Russia} \vskip 24pt

\begin{abstract}
\begin{center}\parbox{14.5cm}
{Difference-phase (or Hanbury Brown - Twiss type) intensity interference of classical light is considered in higher orders in the intensity. It is shown that, while the visibility of sum-phase (NOON-type) interference for classical sources drops with the order of interference, the visibility of difference-phase interference has opposite behavior. For three-photon and four-photon interference of two coherent sources, the visibility can be as high as 81.8\% and 94.4\%, respectively. High-visibility three-photon and four-photon interference of space-time and polarization types has been observed in experiment, for both coherent and pseudo-thermal light.}
\end{center}
\end{abstract}
 \maketitle \narrowtext
\vspace{-10mm}

\section{I. Introduction}

Intensity interference, first discovered in the experiments by Hanbury Brown and Twiss (HBT)~\cite{HBT}, can be defined as the dependence of second- and higher-order intensity moments on certain phase delays, with the intensity itself being independent of these phase delays. Although the Hanbury Brown - Twiss experiment is believed to mark the start of quantum optics, a real burst of interest to intensity interference started with the observation of this effect with two-photon light~\cite{Mandel1}. The remarkable feature of second-order intensity interference with two-photon light is that its visibility can, in principle, be as high as 100\%, while the visibility of second-order intensity interference with classical light cannot exceed 50\%~\cite{Mandel3},~\cite{aspekty}.

One can distinguish between two types of intensity interference, which can be called HBT-type interference and NOON-type interference~\cite{BelKl}. In the HBT-type interference, the second intensity moment depends periodically on the difference of the phases introduced in the output channels, in front of the detectors whose photocount coincidences are registered. A typical example is intensity interference observed for binary stars~\cite{stars}, or two-slit intensity interference of thermal light~\cite{Lugiato},~\cite{Shih},~\cite{Wu}. It is due to this 'phase-difference' dependence that an HBT stellar interferometer is insensitive to atmospheric aberrations. NOON-type interference is observed when the intensity moments depend on the sum of phases~\cite{NOON}. This type of interference has certain outstanding properties provided that the light at the input is in an N-photon state: two-photon state for second-order interference, three-photon state for third-order interference, and so on. The remarkable properties of NOON-type interference are super-resolution~\cite{Dowling} and super-sensitivity to phase measurements~\cite{superphase}. Another important point is that the visibility of NOON-type interference for classical light drops dramatically with the order of interference~\cite{BelKl}. It is this fact that leads to a stronger violation of multi-photon Bell inequalities compared to two-photon ones~\cite{Zukowsky},~\cite{BelKl}.

However, till very recently no consideration was given to multi-photon HBT interference. At the same time, HBT interference can substitute NOON-type interference in important techniques called ghost diffraction and ghost imaging~\cite{Shih+Strekalov},~\cite{Shih+Pittman}. Although observed first with two-photon sources, ghost diffraction and imaging experiments were soon reproduced with classical light~\cite{Lugiato},~\cite{Lugiato2},~\cite{Boyd},~\cite{Shih}. The only disadvantage of classical light with respect to two-photon interference and two-photon ghost imaging, compared to two-photon entangled sources, is the limited visibility, which is always below 100\%. At the same time, in interference and imaging experiments with classical sources the visibility is independent of the intensity, which can therefore be arbitrarily high. In contrast, the available sources of two-photon light (spontaneous parametric down-conversion or spontaneous four-wave mixing) should be sufficiently weak to provide high-visibility interference: while the visibility is close to 100\% for faint two-photon light, it drops with the increase in the mean number of photons per mode~\cite{NL},~\cite{LugiatoPDC}.

In this paper we consider higher-order intensity interference and find, theoretically and experimentally, the classical limit of such interference in the third- and fourth-orders. The obtained values are  81.8\% and 94.4\%, respectively. We show that the visibility of HBT interference for classical light grows rapidly with the order of interference, in contrast to NOON-type interference. This property makes high-order HBT interference of classical sources a good candidate for ghost imaging experiments.

The paper is organized as follows. In Section II we theoretically analyze third- and fourth-order HBT-type interference of two classical light sources and derive the expressions for the maximal visibility achievable for sources with coherent and thermal statistics. Section III describes the experiment on observing third-order interference of two classical sources by means of coincidence counting. Section IV is devoted to the polarization analogue of the observed effect. In Section V, using an alternative method of processing digital images, we register high-visibility third- and fourth-order interference of coherent and thermal light. Finally, conclusions are made in Section VI.

\section{II. Theory}

For our consideration, we chose Young's two-slit geometry~\cite{generality}. This geometry was used in many experiments on two-photon interference and two-photon ghost imaging. This time, however, we consider the interference to be registered in the far-field zone by three detectors instead of two (Fig.1), each detector measuring the instantaneous intensity and triple photocount coincidences being counted by a coincidence circuit. This is the standard experimental technique to measure the third-order Glauber's intensity correlation function (ICF)~\cite{GCF}. Let A, B be point-like classical sources, having the same statistics, the same average intensities, and independently fluctuating phases.

If the fields of the sources are written as $E_{A,B}=E_{0A,B}e^{-i\omega t+i\phi_{A,B}(t)}$  where $E_{0A,B}$  are slowly varying amplitudes (for coherent beams, they are constant) and  $\phi_{A,B}(t)$ are the fluctuating phases, then the instantaneous intensities registered by the detectors are

\begin{equation}
I_n=I_A+I_B+E_{0A}E_{0B}^*
e^{i(\phi_n+\phi_A(t)-\phi_B(t))}+c.c., \label{intens}
\end{equation}
where $I_{A,B}=|E_{0A,B}|^2$ and $\phi_n\equiv\phi_{An}-\phi_{Bn}$,  $\phi_{Sn}$  being the phase accumulated by the radiation of source $S=A,B$  on the way to detector $n=1,2,3$.  Note that $\phi_n$  would determine the phase of the usual (first-order) interference pattern at point $n$ in the absence of random phase fluctuations.

Due to the independently fluctuating phases $\phi_{A,B}(t)$, there is no stable interference pattern in the far-field zone: time averaging makes the intensities in Eq.(\ref{intens}) independent of the phases $\phi_n$. Because of this, transverse displacement of the detectors (shown in Fig.1) does not change the intensities registered by them. However, according to Eq.(\ref{intens}), the instantaneous intensities  at points $n=1,2,3$ are correlated or anti-correlated, depending on the phase difference. Correspondingly, displacement of the detectors will change the intensity correlation functions. Indeed, calculation of the third-order Glauber's correlation function for the intensities at points $1,2,3$ gives

\begin{eqnarray}
G_{123}^{(3)}\equiv\langle I_1I_2I_3\rangle\nonumber
\\
=\langle I_A^3\rangle+\langle I_B^3\rangle+
\left[\langle I_A^2\rangle\langle I_B\rangle+\langle I_B^2\rangle\langle I_A\rangle\right]
\nonumber
\\
\times[3+2(\hbox{cos}\phi_{12}+
\hbox{cos}\phi_{23}+\hbox{cos}\phi_{13})],
\nonumber
\\
\phi_{nm}\equiv\phi_n-\phi_m; n,m=1,2,3.
\label{CF}
\end{eqnarray}
Here, we took into account that due to time averaging, all terms containing the phases $\phi_{A,B}(t)$ turn to zero.

We see that displacement of the detectors influences only the phases $\phi_{nm}$ and has no effect on the intensity moments in (\ref{CF}), which only depend on the statistical properties of the sources $A,B$. This is because we assumed the sources to be point-like; therefore, intensity fluctuations caused by each source do not depend on the transverse coordinates of the points $1,2,3$. In Section III, we will consider the case of extended sources and introduce the corresponding corrections into Eq.(\ref{CF}).

Passing to the normalized third-order correlation function,

\begin{equation}
g_{123}^{(3)}\equiv\frac{G_{123}^{(3)}}{\langle I_1\rangle\langle I_2\rangle\langle I_3\rangle},\langle I_n\rangle=\langle I_A\rangle+\langle I_B\rangle,
\label{norm3}
\end{equation}

we obtain that

\begin{equation}
g_{123}^{(3)}=\frac{g^{(3)}}{4}+
\frac{g^{(2)}}{2}[\frac{3}{2}+\hbox{cos}\phi_{12}+
\hbox{cos}\phi_{23}+\hbox{cos}\phi_{13}],
\label{g3}
\end{equation}
where $g^{(2)}, g^{(3)}$  are, respectively, the second-order and
third-order normalized ICFs for each of the two sources.  Note that not all phases are independent:  $\phi_{13}=\phi_{12}+\phi_{23}$.

The cosine sum in the square brackets of Eq.(\ref{g3}) varies in the range from $-1.5$ to $3$. The maximum corresponds to all $\phi_{ij}=0$ and the minimum, to $\phi_{12}=\phi_{23}=2\pi/3$. To achieve these minimal and maximal values, the phases should vary synchronously, $\phi_{12}=\phi_{23}$. The visibility of the interference pattern will then be given by the expression

\begin{equation}
V^{(3)}=\frac{1}{1+\frac{2g^{(3)}}{9g^{(2)}}}.
\label{V3}
\end{equation}

Because for classical light $g^{(3)}\ge [g^{(2)}]^2$  and  $g^{(2)}\ge 1$~\cite{NL}, with the equality holding only for coherent light, it follows that  $g^{(3)}\ge g^{(2)}$, and it is coherent light that provides the largest visibility,  $V_{coh}^{(3)}=9/11\approx 81.8\%$. Thermal radiation gives a lower visibility, $V_{th}^{(3)}=3/5=60\%$, which is still much higher than the corresponding value in the case of two-photon interference (33\%).

If the number of detectors is four, then one can measure fourth-order normalized correlation function, defined as

\begin{equation}
g_{1234}^{(4)}\equiv\frac{G_{1234}^{(4)}}{\langle I_1\rangle\langle I_2\rangle\langle I_3\rangle\langle I_4\rangle},
\label{norm4}
\end{equation}

Writing the intensities similarly to Eq.~(\ref{intens}) and calculating fourth-order correlations, we get the expression for the normalized fourth-order CF in the form

\begin{eqnarray}
g_{1234}^{(4)}=\frac{g^{(4)}}{8}+
\frac{g^{(3)}}{2}+\frac{3[g^{(2)}]^2}{8}+
\frac{g^{(3)}+[g^{(2)}]^2}{4}[\cos\phi_{12}\nonumber
\\
+\cos\phi_{13}+\cos\phi_{14}+
\cos(\phi_{12}-\phi_{13})+\cos(\phi_{12}-\phi_{14})\nonumber
\\
+ \cos(\phi_{13}-\phi_{14})]+
\frac{[g^{(2)}]^2}{8}[\cos(\phi_{12}+\phi_{13}-\phi_{14})\nonumber
\\
+\cos(\phi_{12}+\phi_{14}-\phi_{13})+\cos(\phi_{13}+\phi_{14}-\phi_{12})].
\label{g4}
\end{eqnarray}

Here, we again took into account that not all phases are independent.

We note that the largest visibility is achieved in the case where the constant term, $A\equiv\frac{1}{8}g^{(4)}+\frac{1}{2}g^{(3)}+\frac{3}{8}[g^{(2)}]^2$, is minimal, and the amplitudes of the two oscillating ones, $B\equiv\frac{1}{4}g^{(3)}+\frac{1}{4}[g^{(2)}]^2$  and  $C\equiv\frac{1}{8}[g^{(2)}]^2$, are maximal. Since for classical light, $g^{(4)}\ge [g^{(3)}]^2/g^{(2)}$, $g^{(3)}\ge [g^{(2)}]^2$ and $g^{(2)}\ge 1$~\cite{NL}, the equalities holding only for coherent sources, the amplitudes A,B, C satisfy the inequalities

\begin{equation}
A\ge \frac{1}{2}+\frac{1}{2}g^{(3)},B\le\frac{1}{2}g^{(3)}, C\le\frac{1}{8}g^{(3)}.
\label{ABC}
\end{equation}

It follows that the maximal visibility corresponds to the case of coherent light, for which  $A=1, B=\frac{1}{2}, C=\frac{1}{8}$.

Analysis of expression (\ref{g4}) shows that its maximum is achieved at all $\phi_{ij}=0$, while the minimum occurs at  $\phi_{12}=\frac{\pi}{2},\phi_{13}=\pi,\phi_{14}=-\frac{\pi}{2}$. The expression for the visibility becomes then

\begin{equation}
V^{(4)}=\frac{1}{1+\frac{g^{(4)}}{8g^{(3)}+9[g^{(2)}]^2}}.
\label{V4}
\end{equation}

The visibility for coherent light is  $V_{coh}^{(4)}=17/18\approx 94.4\%$. For thermal light, $V_{th}^{(4)}=7/9\approx 77.8\%$.

It is interesting to plot the maximal visibility values of this difference-phase interference, together with the maximal visibility values of sum-phase interference~\cite{BelKl}, as functions of the order of interference. Fig.2 shows these dependencies for the case of coherent light. One can see that while the visibility of sum-phase interference decreases dramatically with passing to higher orders, the visibility of difference-phase interference increases.

\section{III. Experiment with coincidence counting.}

In the first series of our experiments, the third-order Glauber's correlation function was measured through the coincidence counting rate of three detectors (Fig.3). As the radiation source, we used a frequency doubled Q-switched Nd:YAG laser with the wavelength 532 nm, pulse duration 5 ns, and the repetition rate 3kHz. Instead of two slits, which should be very precisely matched in width to achieve the maximum visibility of multi-photon interference, a single slit of width 150 $\mu$ was used, followed by a birefringent crystal (calcite). The crystal split the beam into the ordinary one and the extraordinary one; with the slit and the crystal placed between polarization (Glan) prisms oriented at angles $\pm 45^{\circ}$ to the plane of the crystal optic axis, this configuration was equivalent to two identical slits separated by a distance of 1.3 mm.  In the far-field zone, where the interference pattern was formed, the radiation was attenuated using neutral-density filters and fed into a three-arm Hanbury Brown-Twiss interferometer with three photon-counting avalanche photodiodes and a triple-coincidence circuit with the coincidence resolution time 4.2 ns.  Attenuation was necessary to keep the average number of photocounts per pulse much less than one; otherwise, because of the dead-time effect, the photocount and coincidence rates would be measured incorrectly. Due to the gating of the registration electronic system, dark noise was suppressed by several orders of magnitude. In order to scan the interference pattern, plane-parallel glass plates with thicknesses 50 mm and 60 mm, respectively, followed by a 150 $\mu$ pinholes for spatial-mode selection, were placed at the inputs of detectors 1 and 3. Turning the glass plates substituted for moving the detectors, which is usually performed in HBT-type experiments but is more complicated technically. By turning the plates, one could scan the phase of either one or two detectors within the range $0\dots 6\pi$.

In order to study multi-photon interference of sources with thermal statistics, we prepared pseudo-thermal light by means of a rotating ground-glass disc placed after the calcite crystal. With the disc removed, we observed Young's interference of two sources with coherent statistics.

To erase first-order interference in the far-field zone we used an electro-optical modulator (EOM) consisting of four DKDP crystals. The DKDP crystals were oriented with their optic axes parallel to the calcite crystal axis. The electric field applied to the crystals changed their refractive indices and, as a result, the phases between orthogonally polarized components of the light propagating through them. When an AC voltage with the frequency 50 Hz was applied to the EOM the interference fringes in the far-field zone were moving. As a result, first-order interference in the time-averaged intensity distribution vanished. Such a harmonic oscillation of the relative phase between the ordinary and extraordinary beams led to the same effect as random phase fluctuations would: it erased the interference pattern in the time-averaged intensity distribution but did not influence second- and higher-order intensity moments.

Due to the finite size of the slit, the two sources in our setup were not point-like, as it was assumed in the derivation of Eq.(\ref{CF}). In the case of coherent sources, this does not cause any corrections to Eq.(\ref{CF}) since the normalized intensity correlation function of a coherent source is equal to unity everywhere. In the case of pseudo-thermal sources, Eq.(\ref{CF}) has to be modified if the distance between the detectors is comparable with or exceeds the transverse coherence length of the radiation.  The transverse coherence length for each source, $\rho=l\lambda/a$, is determined by the sizes $a$ of the spots formed on the disc by the ordinary and extraordinary beams (0.2 mm),  the distance $l$ from the disc to the detectors and the radiation wavelength $\lambda$. The first-order spatial correlation function of each source A,B in the far-field zone is then

\begin{equation}
g^{(1)}(x)=\left|\frac{\sin (2\pi x/\rho)}{2\pi x/\rho}\right|,
\label{g1}
\end{equation}
and for thermal light, it determines all higher-order correlation functions. Then, in Eq. (\ref{g3}), the normalized ICFs of each of the two sources A,B $g^{(2)}, g^{(3)}$ are not constant but depend on the positions $x_{1,2,3}$ of the detectors in the far-field zone:

\begin{equation}
g^{(2)}\rightarrow g_s^{(2)}(x_i,x_j)\equiv g_{sij}^{(2)}=1+(\gamma_{ij})^2,
\label{Siegert}
\end{equation}
where $\gamma_{ij}\equiv g^{(1)}(x_i-x_j)$,
and

\begin{eqnarray}
g^{(3)}\rightarrow g_s^{(3)}(x_1,x_2,x_3)\equiv g_{s123}^{(3)}
\nonumber\\
=1+(\gamma_{12})^2+
(\gamma_{13})^2+(\gamma_{23})^2+2\gamma_{12}\gamma_{13}\gamma_{23}.
\label{Sieg3}
\end{eqnarray}
Here, the subscript 's' denotes correlation functions of a single source.

Equation (\ref{Siegert}) is known as the Siegert relation, and Eq.(\ref{Sieg3}) is its third-order analogue.

Then, the expression for third-order correlation function in the case of two thermal sources becomes

\begin{eqnarray}
g_{th}^{(3)}(x_1,x_2,x_3)=\frac{1}{4}g_{s123}^{(3)}+\frac{1}{4}(g_{s12}^{(2)}+g_{s13}^{(2)}+g_{s23}^{(2)})
\nonumber\\
+\frac{1}{2}[\hbox{cos}\phi_{12}(\gamma_{12}+\gamma_{23}\gamma_{13})\gamma_{12}
\nonumber\\
+\hbox{cos}\phi_{23}(\gamma_{23}+\gamma_{13}\gamma_{12})\gamma_{23}
\nonumber\\
+\hbox{cos}\phi_{13}(\gamma_{13}+\gamma_{12}\gamma_{23})\gamma_{13}],
\label{ther_g3}
\end{eqnarray}
Note that the phase differences $\phi_{ij}$ are related to the coordinates $x_{i,j}$ as $\phi_{ij}=\frac{2\pi b}{\lambda l}(x_i-x_j)$, where $b$ is the distance between the sources.

Experimental results for the cases of coherent and pseudo-thermal light are shown in Fig.4 a,b, respectively. Points correspond to the measured values of the normalized third-order Glauber's ICF; curves show the fit given by Eqs.(\ref{g3},\ref{ther_g3}), respectively, for the coherent and thermal cases. In accordance with the condition $\phi_{12}=-\phi_{32}$, in our experiment the glass plates in front of detectors 1 and 3 were rotated synchronously, both clockwise (since detector 1 was in the reflected beam and detector 3 in the transmitted beam, this led to the opposite variation of the phases $\phi_{12},\phi_{32}$). The obtained visibility for the case of coherent radiation is 74\%; for the case of pseudo-thermal radiation, 38\%. The small value of visibility in the case of thermal light is due to the finite transverse coherence length of the radiation. In each plot, we also show the spatial dependence of single counts for one of the detectors whose phase was scanned. Although the dependence is not completely flat (the variation is caused by the speckle structure of the laser light and the envelope of the single-slit diffraction pattern), the two-slit interference pattern in the intensity distribution is almost completely erased. Note that the presented ICF was normalized to the product of the three intensities; as a result, the 'noisy' structure of single-photon counts did not influence the third-order interference pattern.

From Eq.(\ref{g3}), we can also find the interference visibility in the case where the third-order interference pattern is scanned by only one of the three detectors. This visibility is maximal if the phase difference for the remaining two detectors is $\pi/2$. The corresponding visibility value is  $V=1/\sqrt{2}$  (approximately 70.5\%), which is lower than in the case of scanning two detectors but still considerably higher than in the case of the second-order interference. Figure 5 shows the results of $g^{(3)}$  measurement with detectors 1 and 2 fixed and detector 3 scanned. The relative phase of detectors 1 and 2 was aligned to be $\pi/2$ using second-order interference pattern for detectors 1 and 2. The observed visibility is 64\%.

\section{IV. Third-order polarization interference.}

The same setup, with some modification, was used to demonstrate the polarization analogue of the observed effect. The modified experimental setup is shown in Fig.6. By means of a quarter-wave plate placed after the EOM instead of the analyzer, the ordinary and extraordinary beams were transformed into right- and left-circularly polarized beams. In the registration part of the setup, a polarizer was inserted in front of each detector. Because of the varying phase between the two beams, light in the far-field zone was not polarized in the first order in the intensity, and rotation of the polarizers did not change the average intensities measured by the detectors.

At the same time, third-order correlation functions in the far-field zone in such a configuration do depend on the mutual orientations of the polarizers. Indeed, consider the sources A and B in Fig.1 to be polarized, respectively, right- and left-circularly, and the detectors 1,2,3 to be preceded by polarizers set at angles $\theta_1, \theta_2, \theta_3$. Then the instantaneous intensity at point $n$ is given by an expression similar to Eq.(\ref{intens}), where the fields   $E_{0A}, E_{0B}$ are substituted by their projections onto polarization state selected by polarizer $n$, $E_{0A}(\bold{e}_R,\bold{e}_n)$, $E_{0B}(\bold{e}_L,\bold{e}_n)$, where the unit polarization vectors can be written in the HV basis as

\begin{eqnarray}
\bold{e}_R=\frac{1}{\sqrt{2}}(\bold{e}_H+i\bold{e}_V), \bold{e}_L=\frac{1}{\sqrt{2}}(\bold{e}_H-i\bold{e}_V), \\
\nonumber
\bold{e}_n=\bold{e}_H\hbox{cos}\theta_n+\bold{e}_V\hbox{sin}\theta_n.
\label{Jones}
\end{eqnarray}

As a result, equations (\ref{CF}), (\ref{g3}), (\ref{g4}) have the same form, with the replacement

\begin{equation}
\phi_{nm}\rightarrow \phi_{nm}+2(\theta_n-\theta_m)
\label{replace}
\end{equation}

If the phases are fixed, third-order and fourth-order intensity moments vary depending on the mutual orientations of the polarizers, although the intensities are constant.

In experiment (Fig.6), we set one of the polarizers at $0^{\circ}$ and rotated the other two polarizers in the opposite directions, with the phases $\phi_n$ set at zero. According to Eqs.(\ref{g3}), (\ref{replace}), third-order normalized correlation function should vary in this case as

\begin{equation}
g_{coh}^{(3)}=1+\frac{1}{2}[\hbox{cos}(2\theta_2)+\hbox{cos}(2\theta_3)+\hbox{cos}(2\theta_2-2\theta_3)]
\label{g3pol}
\end{equation}

Its maximum should be observed at $\theta_2=\theta_3=0^{\circ}$ and its minimum, at $\theta_2=-\theta_3=60^{\circ}$, and the modulation should have 81.8\% visibility. Such a dependence was indeed obtained in experiment (Fig.7). The visibility achieved is 73\%. Solid line shows a fit with Eq.(\ref{g3pol}). The difference between the achieved visibility and the expected one is caused by an inaccuracy in the setting of the phases; for instance, dashed line shows the result of calculation with the phases set at $\phi_{12}=\phi_{32}=\pi/6, \phi_{13}=0$.

\section{V. Experiment with a digital photographic camera.}

For the measurement of fourth-order interference, instead of counting four-fold coincidences, we turned to a different method of measuring spatial ICFs -namely, to processing patterns registered by a digital camera, as suggested in Ref.~\cite{Lugiato_CCD}. The interference pattern in the far-field zone was recorded by a photographic camera Canon Powershot S2 IS. For this experiment, we used a Nd:YAG laser with a repetition rate 47 Hz. The exposure time was 1/50 s, which provided that each frame was made with a single laser pulse. A typical interference pattern recorded in one frame is shown in Fig.8a. To accumulate sufficient statistics, 500 shots were made, both for the coherent case and for the pseudo-thermal case. Due to the phase shifts introduced by the EOM, the phase of the interference pattern varied from frame to frame, so that the intensity spatial distribution averaged over all frames had almost no modulation (the visibility was less than 10\%). The averaging of the intensity distribution and correlation functions was performed over a rectangular spatial area with the dimensions $50$ pixels along $y$ and $600$ pixels along $x$ (shown in Fig.8a). This, in particular, allowed us to eliminate the speckle structure in the intensity distributions (Fig.8b), the interference modulation in which was erased to a considerable extent. The images were processed in the following way: first, each pattern like the one shown in Fig.8a was averaged over $50$ pixels in $y$. This way we obtained one-dimensional patterns $I_j(x), j=1,\dots,500$. These one-dimensional patterns were further processed to obtain the averaged intensity distribution and the correlation functions.

As it was shown above, the maximal visibility of third-order (fourth-order) interference patterns is achieved when two (three) phases $\phi_{nm}$ are varied synchronously. In the triple coincidence counting measurements, this was done by moving two detectors simultaneously. In the method of digital images processing, the same result is achieved by measuring intensity correlations between correctly chosen points of the image. The average intensity and the normalized third- and fourth-order correlation functions were calculated as

\begin{eqnarray}
I(x)=\langle I_j(x)\rangle\equiv\frac{1}{n}\sum_{j=1}^n I_j(x),
\nonumber
\\g^{(3)}(x)=\frac{\langle
I_j(x)I_j(0)I_j(-x)\rangle}{I(x)I(0)I(-x)},\label{calc}
\\g^{(4)}(x)=\frac{\langle
I_j(x)I_j(0)I_j(-x)I_j(-2x)\rangle}{I(x)I(0)I(-x)I(-2x)}. \nonumber
\end{eqnarray}
where the index $j$ numerates the frames, $n$ is the total number of frames ($500$ in our case) and angular brackets denote averaging over the frames.

Fig.9 shows the obtained third-order (a,c) and fourth-order (b,d) interference patterns for coherent (a,b) and pseudo-thermal (c,d) sources. As expected, the distributions in Fig 9a,c are similar to the third-order interference patterns registered by means of coincidence method. The interference visibilities achieved with the photographic camera are 73\% and 59\%, respectively.  The theoretical fit, given by Eq.(\ref{g3}) for Fig.9a and by Eq.(\ref{ther_g3}) for Fig.9c, is shown as a dashed line. Fourth-order interference fringes (Fig.9 b,d) reveal a visibility of 93\% for the coherent case and 81\% for the pseudo-thermal case. The theoretical fit (using Eq.(\ref{g4})) is presented only for Fig.9b, since taking into account the effects of transverse coherence in the fourth-order case leads to too bulky expressions. However, even without the fit in Fig.9d, it is clear that the fourth-order correlation function at the center is anomalously high: the theory predicts a value of $24$, while the experiment gives a nearly twice higher value. The same feature, although less pronounced, is seen in Fig.9c: here, the theoretical value of 6 is exceeded by $25$\%. Second-order interference patterns (not presented here) also reveal slightly excess values of $g^{(2)}$, on the order of $10$\%. These discrepancies between theory and experiment may be explained by the non-stationarity of the pseudo-thermal source, caused by the intensity variation of the scattered light after the rotating disc. 

\section{VI. Conclusion.}

In conclusion, we have considered, both theoretically and experimentally, higher-order HBT-type interference for classical light. We have shown theoretically that in the cases of third-order and fourth-order interference, the largest visibility is achieved for coherent light and this 'classical limit' is 81.8\% in the third-order case and 94.4\% in the fourth-order one. Thermal light provides lower visibility of interference, which is still much higher than in the second-order case and reaches the values of 60\% and 77.7\% for third- and fourth-order interference, respectively. This fact opens interesting perspectives for using thermal light in higher-order ghost imaging experiments~\cite{with LAWu}. Indeed, up to recently thermal light was supposed to be less efficient for ghost imaging than two-photon light, due to a lower visibility. Preliminary calculations show that if one uses high orders in the intensity, this drawback of thermal light is eliminated.

High-visibility HBT interference has been observed in experiment both for coherent and pseudo-thermal light. These results were obtained using two different methods of correlation function measurement: by counting coincidences of several photodetectors and by processing a set of digital images obtained from single laser pulses. High-visibility third-order intensity interference of coherent light has been also demonstrated in polarization measurements.The largest visibility value registered for coherent light was 93\%, for the case of fourth-order interference.

Our results demonstrate a considerable difference in the behavior of the HBT-type interference and NOON-type interference of classical light in higher orders in the intensity. While in the second-order intensity interference, classical light provides not more than 50\% visibility for both these types, the situation in higher orders is different. The visibility of NOON-type interference for classical light is known to reduce rapidly with the order of interference, whereas for HBT-type interference the 'classical visibility limit' grows with the order. Since the low visibility of NOON-type interference for classical light forms the basis for tests of two-photon and higher-order Bell's inequalities, it is very important to distinguish between NOON-type and HBT-type interference in higher orders in the intensity. It also follows from our work that the mere existence of high-visibility interference in the third- and higher-orders in the intensity cannot be considered as a signature of three- or four-photon light. This is in contrast with two-photon interference, where exceeding the 50\% limit is a commonly accepted criterion of nonclassicality.

This work was supported in part by the RFBR grants \#\#06-02-16393, 06-02-39015-GFEN, and the Program of Leading Scientific Schools Support, NSh-4586.2006.2.

\begin{figure}
\includegraphics[height=4cm]{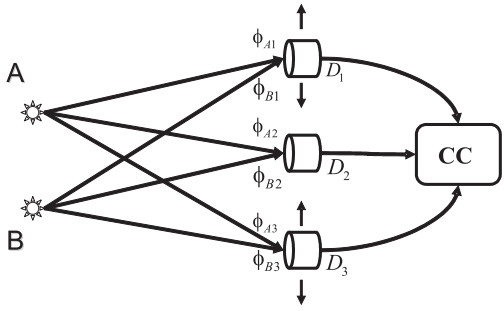} \caption{Young's two-slit interference experiment with the registration of third-order intensity correlations. Radiation produced by two sources A,B is registered by three detectors  $D_1$, $D_2$, $D_3$. Photocounts of the detectors are sent to a coincidence circuit. Transverse displacement of the detectors does not change their photocount rate but changes the triple coincidence counting rate.}
\end{figure}

\begin{figure}
\includegraphics[height=5cm]{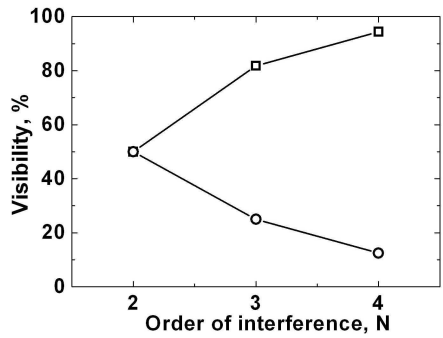} \caption{Maximum visibility for sum-phase (circles)~\cite{BelKl} and difference-phase (squares) intensity interference as a function of the order of interference.}
\end{figure}

\begin{figure}
\includegraphics[height=6cm]{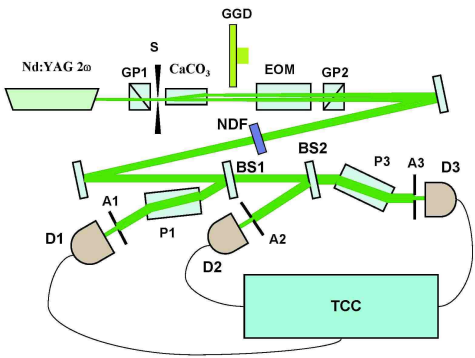}
\caption{Experimental setup. GP1 and GP2, Glan prisms; S, single slit; EOM, electro-optic modulator; BS1 and BS2, beam-splitters;  NDF, neutral density filters; P1  and P3, glass plates; A1-3, apertures; D1-3, avalanche photodiodes; TCC, triple coincidence circuit; GGD,  rotating ground-glass disc.}
\end{figure}

\begin{figure}
\includegraphics[height=9cm]{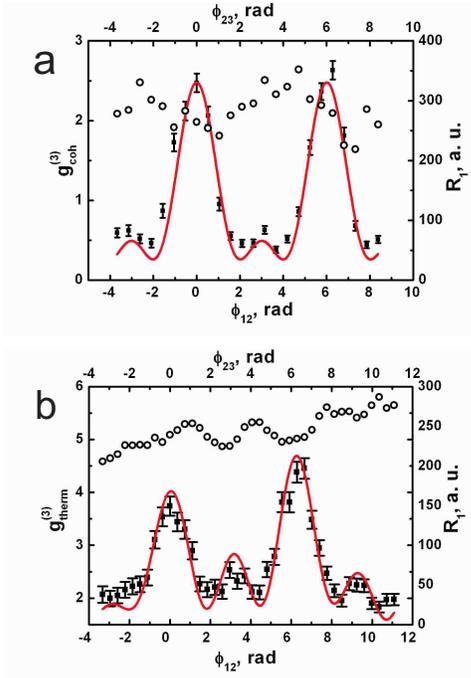}
\caption{Interference pattern in the normalized third-order intensity correlation function for
(a) coherent light and (b) pseudo-thermal light obtained by tilting
glass plates at the inputs of detectors 1 and 3. Empty circles show
the intensity distribution given by the counting rate $R_1$ of
detector 1. Solid lines show the theoretical fit with Eq.(\ref{g3}) (a) and Eq.(\ref{ther_g3}) (b)}
\end{figure}

\begin{figure}
\includegraphics[height=5cm]{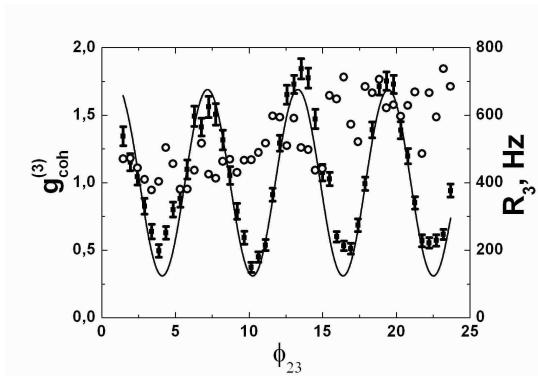}
\caption{Interference pattern in the normalized third-order intensity correlation function for coherent light obtained by tilting the glass plate at the input of detector 3. Relative phase between the first and the second detectors is constant and equal to  $\pi/2$. Empty circles show the intensity distribution (the counting rate R3 of detector 3). Solid line is the theoretical fit with Eq. (\ref{g3}).}
\end{figure}

\begin{figure}
\includegraphics[height=5cm]{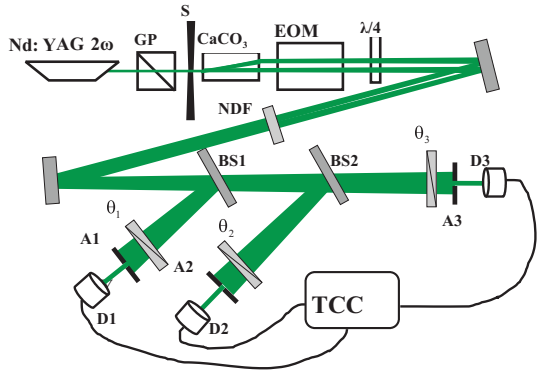}
\caption{Experimental setup for observing third-order polarization interference of classical light. GP, Glan prisms; S, slit; EOM, electro-optic modulator; BS1 and BS2, beam-splitters; $\theta_{1-3}$, orientation of linear polarizers; A1-3, apertures; D1-3, avalanche photodiodes; TCC, triple coincidence circuit; NDF, neutral density filters.}
\end{figure}

\begin{figure}
\includegraphics[height=5cm]{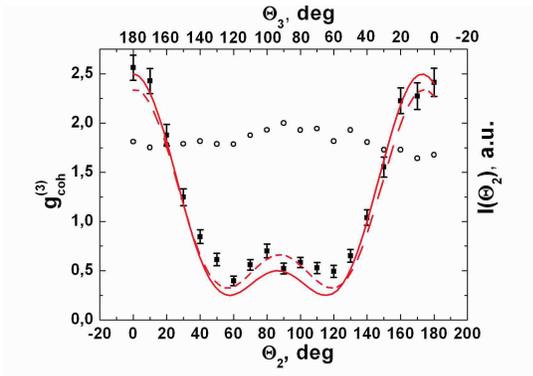}
\caption{Polarization interference in the third order in the intensity observed by synchronously rotating two of the three linear polarizers in opposite directions. Solid line shows the theoretical fit with Eq. (\ref{g3pol}). Dashed line shows the fit taking into account a slight mismatch of the phases, $\phi_{12}=\phi_{32}=\pi/6, \phi_{13}=0$}
\end{figure}.

\begin{figure}
\includegraphics[height=8cm]{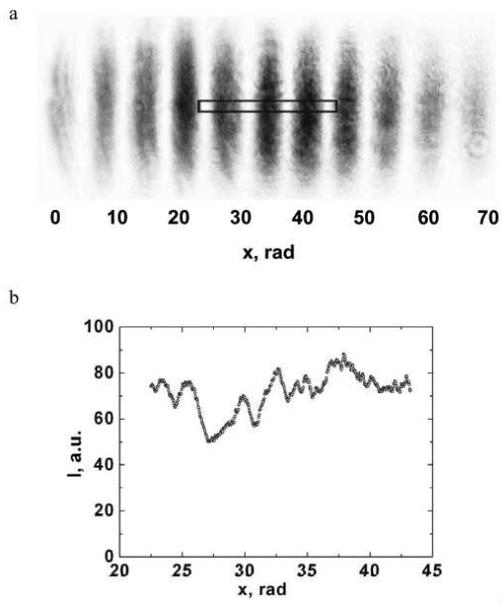}
\caption{(a) Interference pattern recorded in a single laser pulse. The rectangle shows the area over which the analysis was performed: first, the image was averaged over the vertical coordinate (y) and then the resulting one-dimensional distributions were processed according to Eqs.(\ref{calc}).  (b) The intensity distribution I(x) averaged over 500 shots.}
\end{figure}

\begin{figure}
\includegraphics[height=20cm]{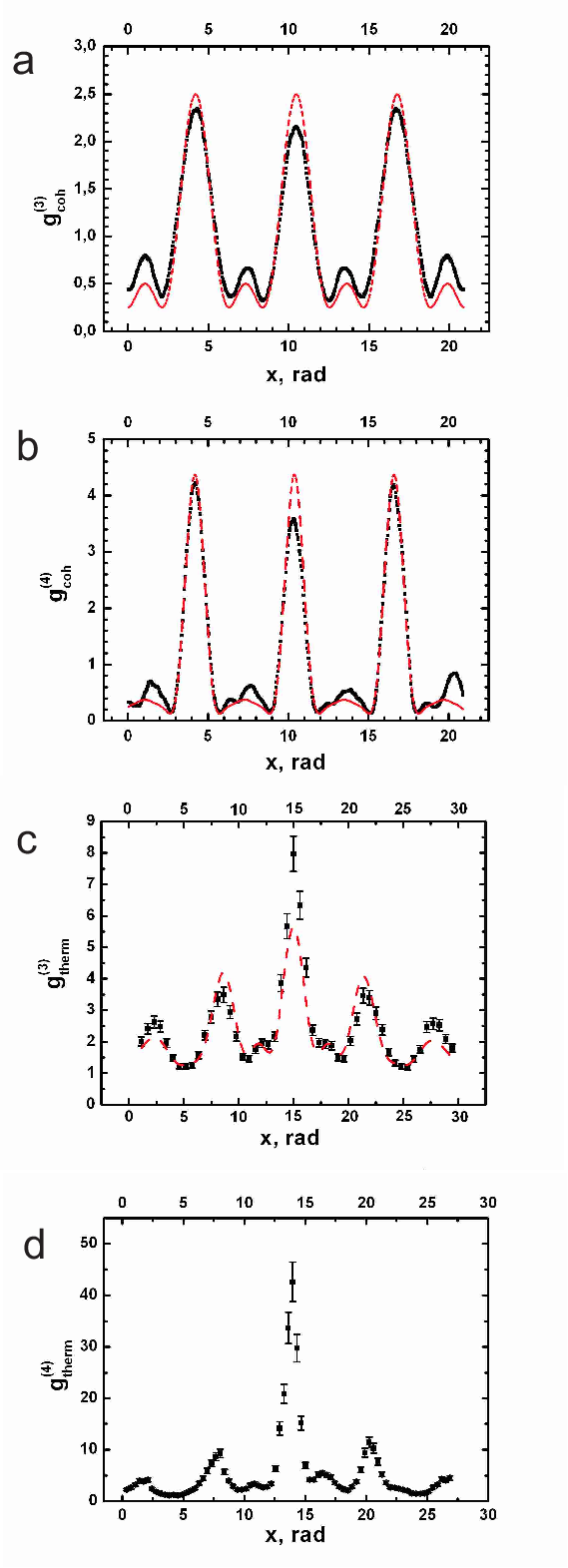}
\caption{Third-order (a,c) and fourth-order (b,d) interference for coherent sources (a,b) and thermal sources (c,d). Dashed line: theoretical fit with Eq.~(\ref{g3}) (a), Eq.~(\ref{g4}) (b) and Eq.~(\ref{ther_g3}) (c).}
\end{figure}


\begin{references}
\bibitem{generality}Description of other types of two-beam
interferometers such as, for instance, Mach-Zehnder one, is similar.
\end{references}
\end{document}